# Super-radiant scattering limit for arrays of subwavelength scatterers

Anna Mikhailovskaya[1,a)] and Konstantin Grotov[2,b)], Dmytro Vovchuk[1], Andrey Machnev[1], Dmitry Dobrykh[1], Roman E. Noskov[1], Konstantin Ladutenko[2], Pavel Belov[2] and Pavel Ginzburg[1]

[1]School of Electrical Engineering, Tel Aviv University, Tel Aviv 69978, Israel
[2]School of Physics and Engineering, ITMO University, St. Petersburg 197101, Russia

[a)] Author to whom correspondence should be addressed: *anna2@mail.tau.ac.il*
[b)] Author to whom correspondence should be addressed: *konstantin.grotov@metalab.ifmo.ru*

**Abstract:**

Electromagnetic scattering bounds on subwavelength structures play an important role in estimating performances of antennas, RFID tags, and other wireless communication devices. An appealing approach to increase a scattering cross-section is accommodating several spectrally overlapping resonances within a structure. However, numerous fundamental and practical restrictions have been found and led to the formulation of Chu-Harrington, Geyi, and other limits, which provide an upper bound to scattering efficiencies. Here we introduce a 2D array of near-field coupled split-ring resonators and optimize its scattering performances with the aid of a genetic algorithm, operating in 19th-dimensional space. Experimental realization of the device is demonstrated to surpass the theoretical single-channel limit by a factor of >2, motivating the development of tighter bounds of scattering performances. A super-radiant criterion is suggested to compare maximal scattering cross-sections versus the single-channel dipolar limit multiplied by the number of elements within the array. This new empirical criterion, which aims on addressing performances of subwavelength arrays formed by near-field coupled elements, was found to be rather accurate in application to the superscatterer, reported here. Furthermore, the super-radiant bound was empirically verified with a Monte-Carlo simulation, collecting statistics on scattering cross sections of a large set of randomly distributed dipoles. The demonstrated flat superscatterer can find use as a passive electromagnetic beacon, making miniature airborne and terrestrial targets to be radar visible.

## I. INTRODUCTION

Scattering cross-section characterizes the interaction between an incident electromagnetic radiation and a body[1]. It is rather convenient to consider the phenomenon in three different regimes, which are defined by comparing an electromagnetic size of a body with a free-space wavelength. Interactions with large objects can be assessed with ray optics tools, wavelength-comparable geometries require performing full-wave analysis, and subwavelength bodies can be addressed with Rayleigh approximation. However, structures made of high dielectric index materials can be both miniature and resonant. Ceramic elements for the MHz-GHz range[2–9] and semiconductor nanoparticles in the optical domain[10–14] are among representative examples. In many practical cases, antenna devices should be tuned to resonance for achieving better transmit and receive performances[15]. Size reduction of devices, operating at low-frequency (kHz-MHz) regimes where implementing wavelength-comparable designs is not practical, is obtained with lumped impedances loading. While in this case, the element can be maintained at a resonance, size reduction implies a significant bandwidth degradation. Chu-Harrington criterion bounds antenna quality factor (Q-factor) form below, relating it to the device form factor, normalized to an operational wavelength[16]. Being formulated to a dipole resonant condition, the limit can be generalized to include higher-order multipoles. Note, that high-Q, being a desirable parameter for strengthening light-matter interaction in the optical domain, has negative implications in antenna design, as it degrades operational bandwidths and has a very negative impact on the channel capacity of a wireless communication channel. Hereinafter, we will concentrate on discussing scattering on subwavelength elements. In this case, multipole expansion is a convenient tool to assess scattering. Each resonant multipole (a scattering channel), being an element of a complete set of basis functions, can contribute to the scattering cross section with $(2\ell+1)\lambda^2/(2\pi)$, where $\ell$ is a total angular momentum and $\lambda$ is a free space wavelength. $3\lambda^2/(2\pi)$ with $\ell =1$ is commonly referred to as a dipolar single-channel limit[17].

To bypass the single-channel limit, several resonant multipoles should contribute constructively to the scattering. In this case, the structure is called a superscatterer[18–26]. It is worth noting that in geometries, lacking a complete rotation symmetry, eigenmodes of a resonator are nontrivially mapped on far-field multipole expansion of scattering[27,28]. Nevertheless, it is quite intuitive that a superscatterer should accommodate multiple resonances at nearly degenerate frequencies. In this case, a significant near-field accumulation in the vicinity of the structure will emerge, making the design to be extremely sensitive to material losses of constitutive elements and fabrication tolerances. Those aspects are well understood in antenna theory in the context of superdirectivity[29]. To cope with those severe, yet solely practical limitations, we have recently introduced structures, based on small wire arrays (wire bundles), pinched into a styrofoam holder. This arrangement is almost lossless at the GHz frequency range and does not require sub-mm accurate positioning of elements in respect to each other[30].

To reduce the effect of near-field accumulation directly on lossy elements, but nevertheless keep it in the interior of the structure, it is quite appealing to investigate designs, made of strongly coupled resonator arrays. The collective response will originate from mode hybridization, which given a proper design, will lead to superscattering performances. Surpassing Chu-Harrington dipolar limit, in this case, will emerge quite straightforwardly. The challenge, however, is to formulate a tighter upper bound on the scattering cross-section. Here we will phenomenologically introduce a super-radiant scattering limit, which makes an assessment of structures, made of near-field coupled resonators. In this case, the scattering cross-section will be compared with a single-channel dipolar limit multiplied by the number of elements within the array. In other words, can the coupling improve the scattering performances? It is worth noting that related assessments were done across different disciplines, e.g.[31,32]. Based on our recent investigations and the current report, it will become evident that surpassing this new limit is quite challenging, if even possible. An appropriate terminology here is a "super-radiant scattering limit", as the phenomenon shares similarities with the quantum effect of superradiance[33]. The essence of the latter is the acceleration of the spontaneous decay rate from N quantum systems owing to their mutual phase-locking. In our case, the assessment of the total scattering cross-section $\sigma_{tot}$ will be as follows:

$$\sigma_{tot} > \sum_{i=1}^{N} \sigma_i, \qquad (1)$$

where $\sigma_i$ is the scattering cross-section of each individual element within the array.

Here we will investigate structures, based on near-field coupled split-ring resonators (SRRs). Subwavelength SRRs support resonant magnetic dipole modes at GHz spectral range and do not require additional loading with lumped elements[34–37]. Having enough degrees of freedom to tune electromagnetic parameters, SRRs are promising candidates for superscattering designs.

The manuscript is organized as follows: the super-radiant scattering limit is investigated with the aid of discrete dipole approximation, motivating the further development and optimization of structures. Analysis of a single element (SRR) performance and a brief assessment of small arrays is done next. Scattering cross-sections of the structures are normalized to the number of elements within the arrays to find an optimal number of elements for further investigations. Given this size (6 in our case), a genetic algorithm is set up to optimize the structure furthermore. Experimental assessment of the final design comes next. Discussions on the scattering cross-section bounds come before the conclusions.

## II. THE SUPER-RADIANT LIMIT

To assess the new bound, we will consider several mutually interacting point scatterers, applying coupled dipoles formalism [38]. A dipole moment ($\vec{p}(\vec{r})$) is proportional to the local electric field ($\vec{E}_{loc}$) and particle's polarizability:

$$\vec{p}(\vec{r}) = \varepsilon_0 \overleftrightarrow{\alpha} \vec{E}_{loc}(\vec{r}), \qquad (2)$$

where $\overleftrightarrow{\alpha}$ is a polarizability tensor and $\varepsilon_0$ vacuum permittivity. For the sake of simplicity, magnetic and magneto-electric interactions are ignored[39,40]. $N$ dipoles problem is than formulated within $3N$ linear equations, taking into account the vectorial nature of the problem:

$$\vec{p}_i(\vec{r}_i) = \varepsilon_0 \overleftrightarrow{\alpha}_i \left[ \vec{E}_0(\vec{r}_i) + \sum_{n \neq i}^{N} \overleftrightarrow{G}(r_i, r_n) \vec{p}_n(\vec{r}_n) \right], i = 1..N, \qquad (3)$$

where $\overleftrightarrow{G}$ is the Green tensor of the single dipole and $\vec{E}_0$ is the incident field. This set of equations can be solved by a matrix inversion, which allows calculating dipole moments self-consistently. Then, the extinction cross section can be obtained from the optical theorem:

$$C_{ext} = \frac{4\pi k}{|\vec{E}_0|^2} \sum_{i=1}^{N} Im(\vec{E}_0^*(\vec{r}_i) \cdot \vec{p}_i(\vec{r}_i)), \qquad (4)$$

where $k = 2\pi/\lambda$ is the wave number and the incident field amplitude variation on the array is neglected (plane wave excitation is assumed).

For assessing the super-radiant limit, the following numerical experiment will be performed: $N$ resonant point dipoles are randomly distributed in a subwavelength cubic volume with the side of $\lambda/5$. The dipoles are not allowed to approach each other by a distance, smaller than $\lambda/40$. The dipoles are isotropic, lossless and identical with polarizability of a subwavelength sphere, made of a lossless Drude material. Relative permittivity ($\varepsilon_r = -2$) is the resonance condition. Radiation corrections are included in the polarizability model[41]. This numerical setup is rather arbitrary and solely serves for assessing possible extinction cross sections statistically.

Figure 1 shows the probability distributions of the normalized cross sections for several dipoles in the box - $N$, indicated in captions. The horizontal axis is $C_{ext}^{total}/N \cdot C_{ext}^{single}$, where $C_{ext}^{single}$ is the maximal extinction of a single resonant dipole. $C_{ext}^{total}$ is the maximal extinction of the array. Note, that the resonance of the coupled system and the single dipole can be shifted in frequency. The probability distribution is calculated upon assessing 1000 random realizations (uniform distribution in the volume, no correlation between the dipole locations). The super-radiant limit is 1 on the horizontal axis. Panels (a) and (b) differ by dipoles polarizability. 'Strong' correspond to the lossless' dipole polarizability at its resonance, 'weak' to the same value, divided by 10. The following observations can be made: (i) no one of the realizations broke the limit statistically, (ii) increasing the number of dipoles in the volume shifts the distribution to the left, i.e., moving away from the limit (on average). It is worth noting that the local field within the array is extremely non-uniform, making effective medium theories[42] barely applicable. Insets, to the right of the probability distributions demonstrate realization from the sample space. The realizations correspond to the maximal, mean, and minimal scattering, which were observed. It is worth noting that the realizations have no pronounced spatial arrangement and, as the result, are hardly predictable. It is quite obvious that probability distributions in the 'weak' polarizability regime, implying minor dipole-dipole coupling demonstrate, are peaked in the vicinity of 1. Increasing the number of dipoles in the box (the inset) results in the reduction of the scattering cross section on average. Note, that the overall scattering cross-section in the 'strong' polarizability regime are higher than in the 'weak' case, which increases the challenge in finding a proper configuration.

This analysis demonstrates that approaching the limit with a large number of dipoles without an extensive optimization is challenging. Furthermore, it is impossible to draw a conclusion on whenever the limit can be overcome. The objective of the next sections is to assess this limit from the practical standpoint. Magnetic dipoles instead of electric ones will be used due to several practical aspects.

## Strong

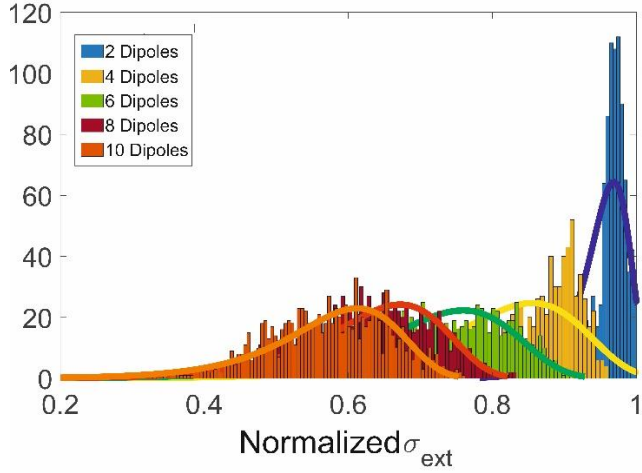
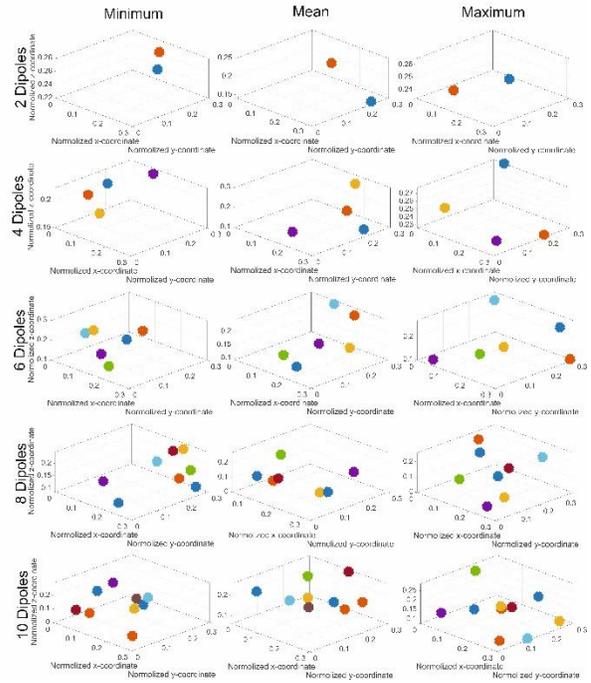

## Weak

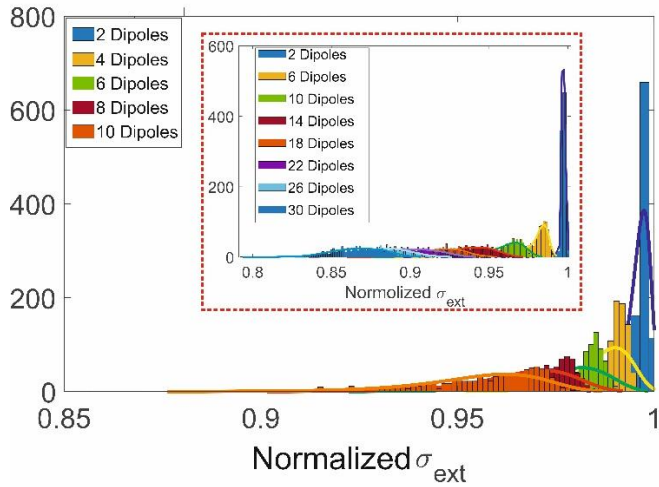
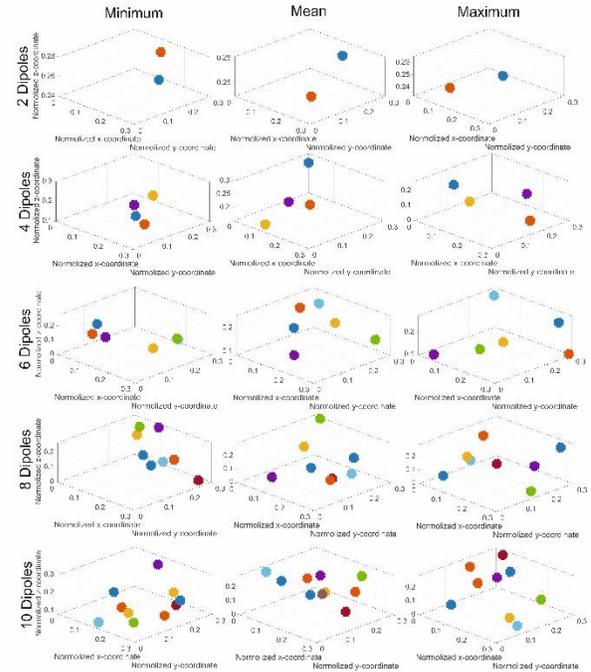

FIG 1. Assessment of the super-radiant scattering limit - probability distribution on the normalized extinction cross-section ($C_{ext}^{total}/N \cdot C_{ext}^{single}$) for $N$ coupled dipoles in a subwavelength volume. (a) Strong polarizability case – resonant lossless dipoles. (b) Weak

polarizability – 1/10 of the previous case. Right insets – minimum, mean, and maximum realizations of the dipoles arrangement from the sample space.

III. THE ELECTROMAGNETIC DESIGN AND STRUCTURE'S OPTIMIZATION

Prior to analyzing arrays, performance of single elements will be briefly surveyed. Electromagnetic modelling was performed in CST Microwave Studio Suite, Frequency Domain Solver. The number of mesh cells was approximately $1.5-2\times10^5$ for all models. For basic element we chose double circle split ring resonator (CSRR), as it has a smaller footprint, higher Q-factor, and symmetrized response – all in respect to a single SRR. CSRR can be tuned to a resonance at GHz-spectral range without a need to introduce additional lumped impedances. CSRRs are implemented by etching copper strips on a dielectric substrate (Isola IS680 AG338, $\varepsilon_r \approx 3.338$, $tan(\delta) = 0.0026$) – relative permittivity and loss tangent). The inner and outer radii of the ring are $r_{in}=1.9$ mm and $r_{out}=3.5$ mm respectively. The metal strips width is *1 mm* and thickness *35 μm*, with the gap between the inner and outer rings of *0.4 mm*. The upper split in the outer ring and the symmetric split in the inner ring have the same width of *1 mm*. Single CSRR on a substrate was tuned to resonate at 5.2 GHz (magnetic dipole predominates the interaction), with the Q-factor of 61.

At the next stage the number of elements within the array will be chosen. Figure 2(b) is a color map, showing the total normalized scattering cross-section as a function of frequency and a number of elements within the array (2-10 element arrays were investigated). As the last variable is discrete, a linear interpolation has been made. Horizontal cuts correspond to the normalized scattering spectra. The normalization is made by dividing the values by the scattering cross-section of a single CSRR - $\sigma_{tot}/max(\sigma_{tot}^{Single\ CSRR})$. In all the cases, the elements were equally distributed on a circle with $r = 25$ mm radius. This number is quite arbitrary, nevertheless it was chosen to keep the structure electrically small, e.g., $2r/\lambda < 1$. In those studies, all the CSRRs were mutually aligned (their gaps were kept parallel, as it appears in Fig. 2(a)). While single channel dipolar ($\ell=1$) limit for an ideal CSRR is $3\lambda^2/(2\pi) \approx 15.9$ cm$^2$, in the presence of a lossy substrate the practically achievable value is $\approx 11$ cm² (at 5.2 GHz). The colormap in Fig. 2(b) clearly indicates that 6-element array is the best candidate for further investigating superscattering effect. $\sigma_{tot}/max(\sigma_{tot}^{Single\ CSRR})$ is approaches maximum for 6 CSRRs and almost saturates for larger arrays. Relying on this observation, we will concentrate on this 'magic' number hereafter, nevertheless, any other can be chosen as a starting point. Note, that the saturation effect has nothing to do with statistical distributions in Fig. 1, as it relates to specific pre-optimized realizations.

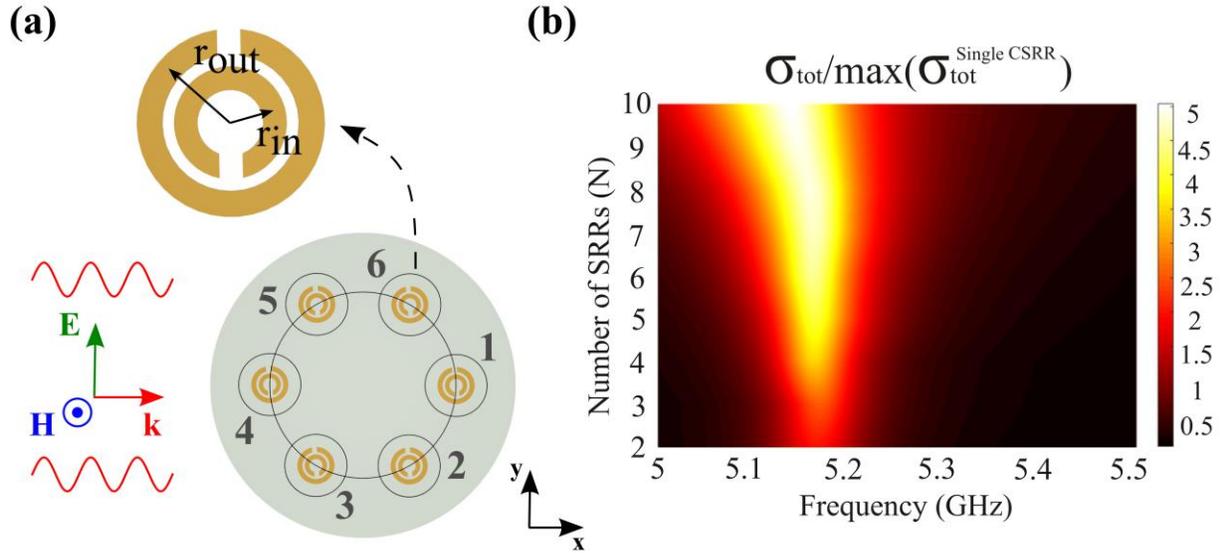

FIG 2. (a) Schematics of a single CSRR, array, and the incident plane wave. (b) Numerical analysis - the total scattering cross-section, normalized to the maximal total scattering of a single CSRR, as a function of frequency and the number of elements within the array.

While the previously investigated structure is rather symmetric, symmetry breaking can lead to a significant improvement of certain electromagnetic parameters. A good historical example is photonic crystal cavities, where a miniature displacement of holes around a defect was found to boost Q-factors by orders of magnitude[43,44].

Scattering cross section maximization is an optimization problem in a multiple variables space. A choice of an optimization algorithm is always a subject to several tradeoffs, compromising between computational efforts and reliability of a solution. Evolutionary algorithm (particle swarm optimization will be used here) starts its initial iteration with a set of random vectors (individuals). There are well-understood conditions, where this algorithm converges to a global extremum with a probability of 1 [45]. Nevertheless, it is quite difficult to fulfill all the requirements to ensure the convergence, which than comes at an expense of computational efforts. Consequently, in many practical applications, the possible heuristic proof of a global minimum is waived in a favor of resource reduction[46,47]. Inverse design using deep learning is another approach, which can be employed for the optimization (e.g. [48,49]). However, machine learning models, e.g. deep neural networks, require a large amount of training data and are less useful for designs, discussed here.

Evolutionary algorithms apply basic provisions, adopted from biological evolution theory. Main steps consist of selection, mutation, and crossover. A random initial population evolves in accordance with the selection rules (which depend on the fitness function) and only the best individuals reach the next iteration of the algorithm. Being first considered at 1956[50], evolutionary methods are extensively used nowadays in various fields of physics. Antenna design[51,52], including the famous NASA evolved antenna[53], 2D materials design[54–56], development of reflective and absorbing structures[57] and metasurfaces[58,59], design of artificial magnetic metamaterials[60] and plasmonic nanoparticles[61] are among the examples. Hereafter, we will use a particle swarm optimization algorithm, where the following parameters are chosen to form the search space: the radius of the circle $R$ on which the CSRRs are initially located, the angle of rotation $\alpha_i$ of each CSRR relative to its center, and the position of the element in a neighborhood of the starting point (this parameter was chosen as $\delta = R/3$ to prevent a potential geometrical overlap between the array elements). For convenience, the position of the $i^{th}$ ring is given in polar coordinates ($\rho_i$, $\phi_i$). Thus, the optimization vector of parameters contains 19 components. The radius ($R$) is varied from 10 mm to 60 mm, the rotation angles ($\alpha_i$) of each CSRR from $-\pi$ to $\pi$, $\rho_i$ from 0 to 5 mm and $\phi_i$ from $-\pi$ to $\pi$. Zero angle corresponds to y- axis direction. The final set of the parameters, obtained with the optimizer, are summarized in Table 1, Supplementary.

Figure 3 is a schematic representation the optimization algorithm flow. At the beginning, a population of N individuals (random vectors of parameters) is formed. Each of the individuals corresponds to a certain design, which was modeled in CST Microwave Studio. The excitation is kept the same – a plane wave propagates along x-axis with the magnetic field polarized along z-axis. The maximal scattering cross-section was constrained to appear at 5 - 6 GHz interval. Setting a hard restriction on the resonant frequency might cause conversion issues and, hence, this parameter is better to be loosely defined. After an integration with the direct solver (evaluation of the electromagnetic problem with CST), the algorithm singles out optimum individuals of the population, which proceeded to the stage of crossing and mutation. As a result of the crossover, a new generation of individuals is created. Further, the mutation operator is applied to the resulting new generation, the purpose of which is to add a small perturbation to the components of the vectors of the new population (the implementation of the mutation operator may also differ from the implementation of the algorithm). At the end of each step of the algorithm, the scattering spectra of new individuals are calculated, and new best representatives of the population are selected. This process is repeated $N_{iterations}$ times and after exiting the algorithm, the best individual is obtained - the design that corresponds to the best-found scattering spectrum.

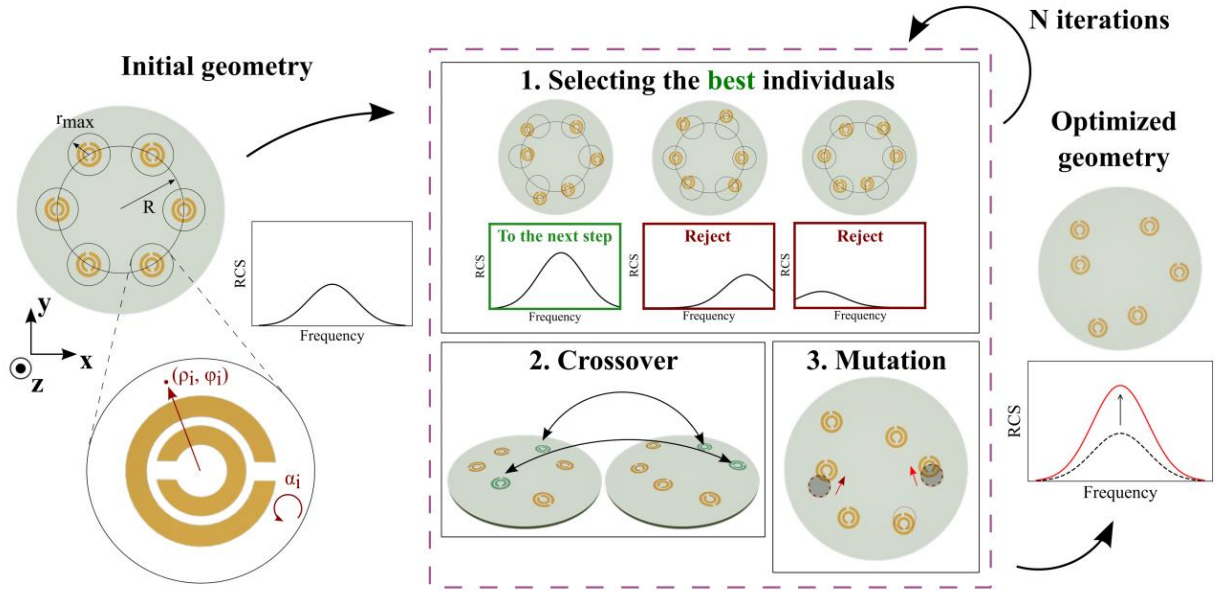

FIG 3. Scheme of the optimization algorithm.

The numerical experiment was carried out with $N_{iterations}$ = 1000. The runtime on 512 GB RAM 3.3 GHz, is 2000 minutes (~2 minutes for a single iteration). The same machine was used to run the algorithm and CST.

Our final superscatterer design is shown in Fig. 4(a). Parameters of the structure are summarized in Table 1 (Supplementary). Fig. 4(b) shows the comparison of scattering cross-sections of a single CSRR, unoptimized array, and the result of the genetic algorithm. The growth of the scattering peak can be clearly seen with genetic design prevailing the unoptimized counterpart by the factor of 1.4. The resonant frequencies of all those 3 structures differ slightly from each other, as the result of relaxing this variable in the optimization.

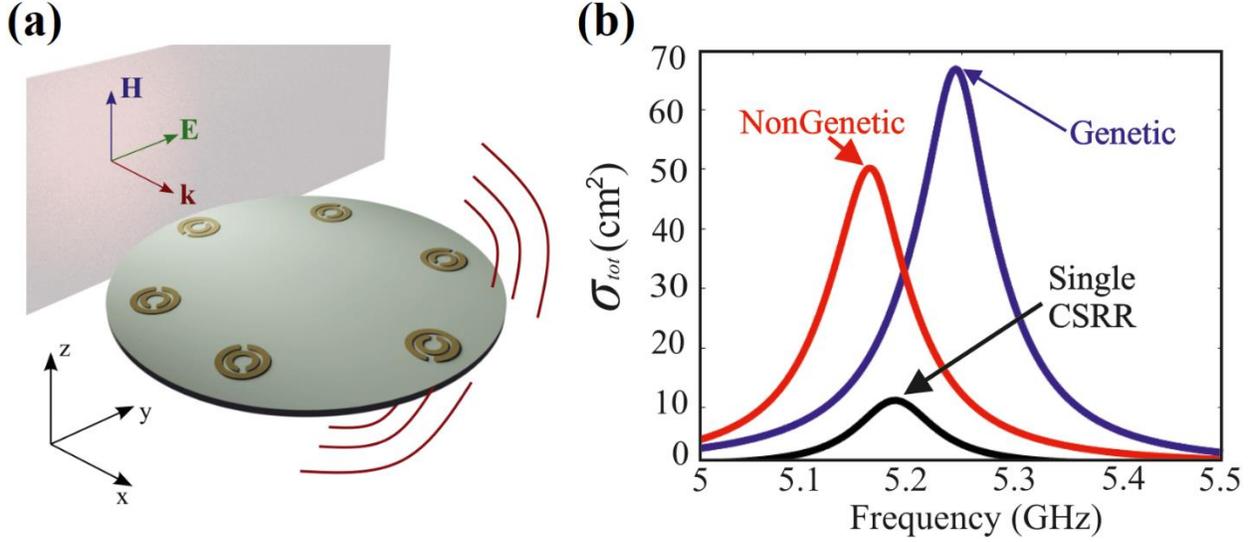

FIG 4. (a) Schematics of a flat superscatterer - 6 CSRRs are distributed on a substrate (Isola IS680 AG338). Illumination is a plane wave, propagating along the X axis and polarized along the Y axis. (b) Scattering cross-section spectra of the superscatterer (genetic design) - blue solid line, equidistantly distributed 6 SRRs on circle - red solid line, and a single CSRR - black solid line.

To reveal the device operation, multipolar expansion of the scattering spectra should be done. In this case, finite element method (COMSOL Multiphysics) was used. The multipole expansion of the scattering cross-section, i.e., the sum of the contributions from different multipole moments up to the third order is given by formula for Cartesian multipoles [50–52]:

$$\sigma_{sca}^{tot} = \sigma_{sca}^{p} + \sigma_{sca}^{m} + \sigma_{sca}^{Q^e} + \sigma_{sca}^{Q^m} + \sigma_{sca}^{O^e} + \sigma_{sca}^{O^m} \approx \frac{k^4}{6\pi\varepsilon_0^2|E_0^2|}|p_j|^2 + \frac{k^4\varepsilon_h}{6\pi\varepsilon_0^2c^2|E_0^2|}|m_j|^2 +$$

$$\frac{k^6}{80\pi\varepsilon_0^2|E_0^2|}|Q_{jk}^e|^2 + \frac{k^6\varepsilon_h^2}{80\pi\varepsilon_0^2c^2|E_0^2|}|Q_{jk}^m|^2 + \frac{k^8\varepsilon_h^2}{1890\pi\varepsilon_0^2|E_0^2|}|O_{jkl}^e|^2 + \frac{k^8\varepsilon_h^3}{1890\pi\varepsilon_0^2c^2|E_0^2|}|O_{jkl}^m|^2, \quad (5)$$

where $|E_0|$ is the electric field amplitude of the incident plane wave, $k$ is the wavenumber, and $c$ is the speed of light, $\varepsilon_h$ is the permittivity of the host media (air in our case), $\varepsilon_0$ is the permittivity of vacuum, $p_j$ and $m_j$ are the electric (ED) and magnetic dipole moments (MD), $(Q_{jk}^e)$ and $(Q_{jk}^m)$ are the electric and magnetic quadrupoles (EQ and MQ), $(O_{jkl}^e)$ and $(O_{jkl}^m)$ electric and magnetic octupoles (EO and MO).

Figure 5 summarizes the expansion results for the non-optimized array and for the array obtained by the genetic algorithm. Recall, that the non-optimized array was nevertheless tuned to its resonance, exhibiting high scattering performances. While in both of the cases, multipole contributions have overlapping resonances, the genetic design leads to a better collocation and, remarkably, brings a MD resonance, which is missing in the initial array (dashed brown line in Fig. 5). Overall, the multipole series reproduce the peak, nevertheless, the exact conversion is not obtained. The conclusion here is that higher order multipoles in the series are missing. Including them explicitly is a rather complex task, as mathematic formulation becomes involved. It is worth noting that the lack of conversions in electrically small structures is extremely rare and, typically, several multipoles describe the interaction quite accurately. Constructive interference of 6 multipoles in our structure become evident.

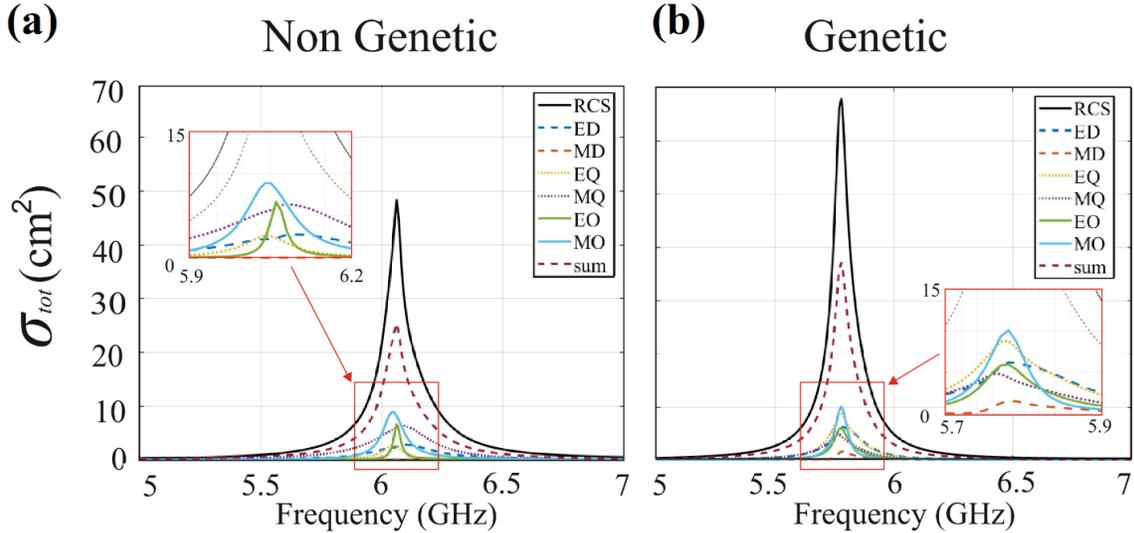

FIG 5. Multipole expansion of the scattering cross-section of (a) a non-optimized array and (b) array, obtained with the genetic algorithm. Abbreviations appear in captions and elaborated in the main text.

## IV. EXPERIMENTAL MEASUREMENTS

The sample, consisting of 6 CSRRs was manufacturing by chemical etching. Isola IS680 AG338 ($\varepsilon_r$ = *3.338, tan($\delta$)* = *0.0026*) was used as a low-loss substrate. The thickness of the dielectric support is *0.2 mm*, while the thickness of the copper layer is around *35 μm*. Those parameters were used to reduce the influence of the substrate on the array's performance. The CSRRs are arranged according to the layout, provided by the algorithm. The fabrication process was optimized to provide high-quality samples with sub-mm precision in all the parameters, though, without using clean room facilities. The experimental sample is shown in the inset to Fig.6(b).

Experimental spectra are shown in Fig. 6(b). Several angles (see captions) of incidence were considered. The sample was rotated around its axis (see Fig. 6(a)), while the magnetic field was always polarized along the CSRRs normal. Optical theorem was used to evaluate the total scattering cross-section. The angular dependence here is relatively weak, making the device attractive from an applied point of view, as an accurate alignment is not required. Black dashed line shows the scattering spectrum of a single CSRR. Its peak is 5 times smaller than the maximal scattering cross-section of the array.

## V. DISCUSSION

After demonstrating the structure's performances, assessing fundamental limits made possible.

**(i) Single-channel limit**

The single-channel limit was defined in the introduction. For this structure it is 17.9cm$^2$ (at 4.9 GHz). Note, that the single CSRR with the maximal scattering of 8.146 cm$^2$ does not reach this limit because of losses. Our structure was found to beat the single channel limit by $\approx 2.17$.

**(ii) Super-radiant scattering limit**

Our newly introduced limit will be assessed next, considering $\sigma_{tot}^{structure}/(N\sigma_{tot}^{single\ CSRR})$, where $N = 6$. This assessment allows underlining the impact of near field coupling on the total scattering. Fig. 6(c) demonstrates the limit for different angles of incidence. The maximum is obtained for 0°, for which the structure was initially optimized. It is worth noting that the numerical analysis predicts overcoming the super-radiant limit by a small fraction, while the experimental results drop below this value. This is rather good indication that the empirical formulation of this new bound is rather tight and difficult to bypass. The experimental sample is rather sensitive to many factors including substrate losses, fabrication accuracy, and the surrounding environment. While the chemical etching was made quite accurate, the substrate permittivity might have fluctuations as well as the copper layer. All those aspects lead to deviations between the theoretical predictions and practical results.

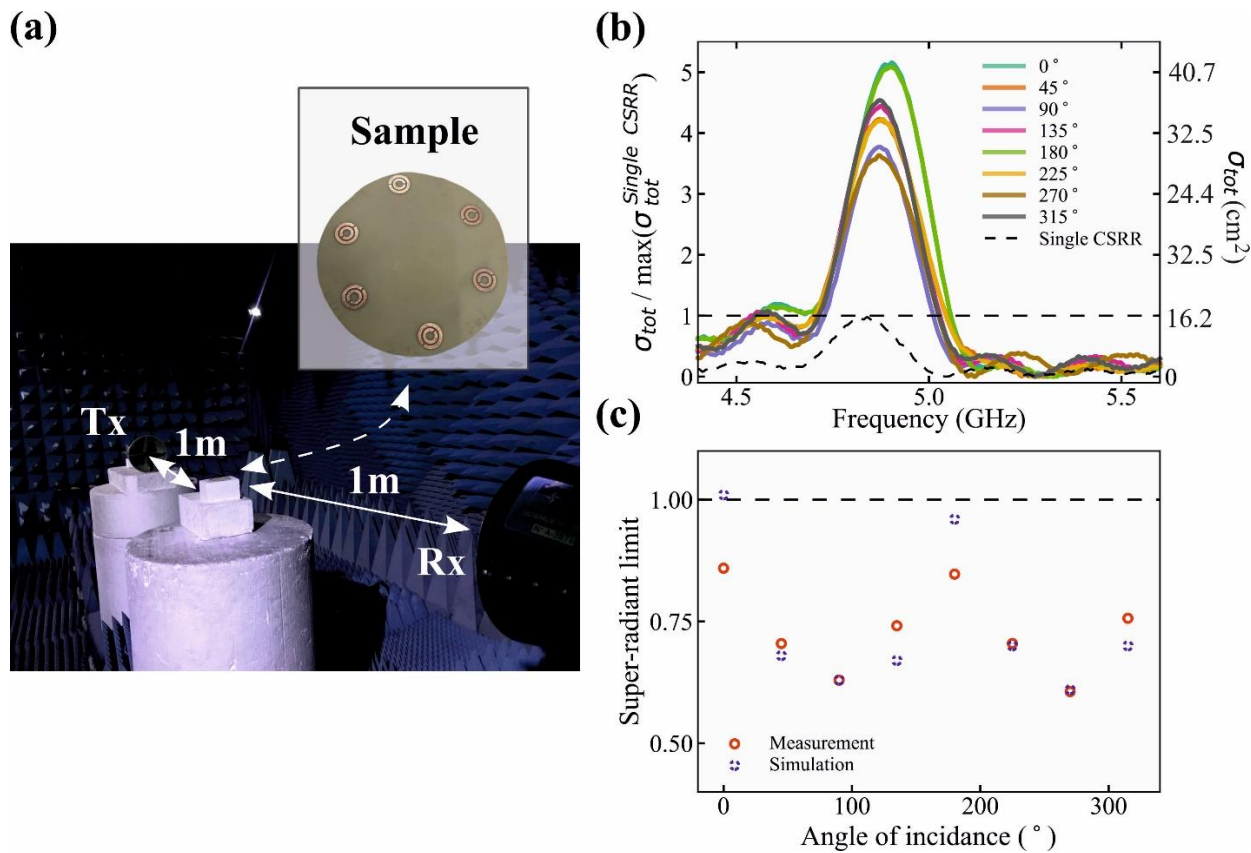

FIG. 6. (a) Experimental setup to measure scattering cross-section. Inset – the sample. (b) Experimental total scattering cross-section spectra for different angles of incidence (in captions). (c) Super-radiant limit, experimental and numerical data. Different angles of incidence are considered.

## VI. CONCLUSION

The design and experimental realization of the superscatterer, based on split-ring resonators array was demonstrated. Genetic algorithm, optimizing 19 independent degrees of freedom, has been implemented to design spectral overlap of 8 multipolar contributions at the same frequency. The experimental sample was shown to surpass the single-channel limit by a factor of 2.17 and motivated the development of new more practical bounds to assess scattering performances of structures, made of near-field coupled elements. A new super-radiant bound has been formulated, suggesting a comparison of the maximal scattering cross-section with a single channel dipolar limit multiplied by the number of elements within the array. The bound was empirically assessed with the aid of Monte-Carlo simulation. We performed a numerical experiment, calculating scattering cross sections of a large set of randomly distributed point dipoles, placed within a subwavelength volume. Any realization succeeded to overcome the super-radiant bound, suggesting its accuracy, yet on a statistical basis. While quite a few physical limits in electromagnetism have been developed[65], the super-radiant limit is quite appealing owing to its applied simplicity.

The designed 6-element array succeeded for overcome this new bound by a small fraction in theory, while the experimental values were found below the limit. Those results indicate that this new assessment sets a rather tight limitation and promote its further use in the field of superscattering.

The demonstrated superscatterer is two-dimensional and it is implemented on a thin flexible lightweight substrate. Similar designs can find a use in many practical applications, including electromagnetic passive beacons, alignment marks for indoor navigation, radar chaff and many others.

ACKNOWLEDGMENTS

TAU team was supported by the Department of the Navy, Office of Naval Research Global under ONRG award number N62909-21-1-2038. There is no joint funding between the collaborating teams.